\documentclass[twocolumn,prb]{revtex4-1}
\usepackage{amsmath}
\usepackage{graphicx}
\usepackage{amssymb}

\begin{document}

\title{Radial Spin Helix in Two-Dimensional\\ Electron Systems
with Rashba Spin-Orbit Coupling}

\author{Yuriy V. Pershin}
\email{pershin@physics.sc.edu}

\affiliation{Department of Physics and Astronomy and USC
Nanocenter, University of South Carolina, Columbia, SC 29208, USA}

\author{Valeriy A. Slipko}
\affiliation{
Department of Physics and Technology,
V. N. Karazin Kharkov National University, Kharkov, Ukraine}

\begin{abstract}
We suggest a long-lived spin polarization structure, a radial spin
helix, and study its relaxation dynamics. For this purpose,
starting with a simple and physically clear consideration of spin
transport, we derive a system of equations for spin polarization
density and find its general solution in the axially symmetric
case. It is demonstrated that the radial spin helix of a certain period relaxes slower than homogeneous spin polarization and plain spin helix.
Importantly, the spin polarization at the center of the radial
spin helix stays almost unchanged at short times. At longer times,
when the initial non-exponential relaxation region ends, the
relaxation of the radial spin helix occurs with  the
same time constant as that describing the relaxation of the plain
spin helix.
\end{abstract}

 \maketitle

\section{Introduction}

At the present time, there is a significant interest in the field
of electron spin relaxation in semiconductors stimulated by
possible future applications of spins in electronics and computing
\cite{Awschalom02a,Zutic04a}. In many two-dimensional (2D)
electron systems the leading  mechanism of spin relaxation is the
D'yakonov-Perel' spin relaxation mechanism
\cite{Dyakonov72a,Dyakonov86a}. Within this mechanism, electron
spins feel an effective momentum-dependent magnetic field
randomized by electron scattering events resulting in relaxation
of electron spin polarization. A number of theoretical and
experimental studies on peculiarities of D'yakonov-Perel' spin
relaxation were reported in the last decade
\cite{Kiselev00a,Sherman03a,Weng04a,Pershin04a,Pershin05a,Jiang05,Bernevig06a,Schwab06a,Koralek09a,Kleinert09a,Duckheim09a,Tokatly10a}.

It was shown in Ref. \onlinecite{Pershin05a} that the spin
relaxation time for 2D electrons depends not only on material
parameters (e.g., strength of spin-orbit interaction, electron
mean free path, etc.) but also on the initial spin polarization
profile. In particular, it was demonstrated that a {\it plain spin
helix} in a 2D electron system with Rashba spin-orbit interaction
 has a longer spin relaxation time than a
homogeneous spin polarization \cite{Pershin05a}. A later study
\cite{Bernevig06a} revealed that in a system with both Rashba
\cite{Bychkov84a} and Dresselhaus \cite{Dresselhaus55a}
interactions such an increase in spin relaxation time can be even
more dramatic. This effect was also observed experimentally
\cite{Weber07a,Koralek09a}.

\begin{figure}[b]
\centering
\includegraphics[width=7cm]{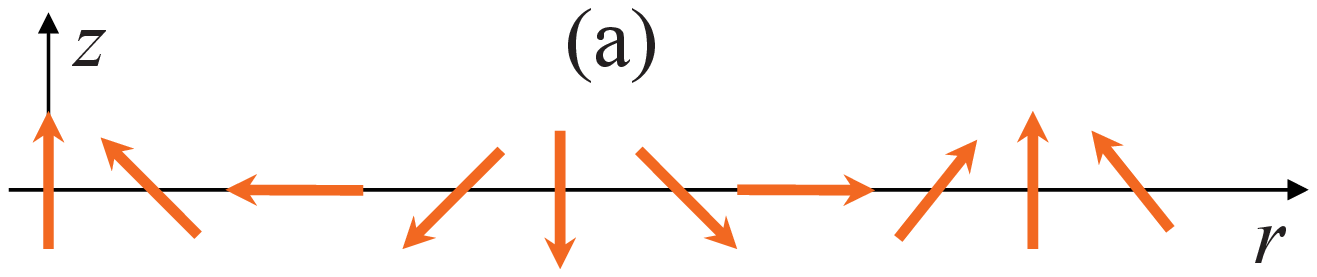}
\includegraphics[width=7.5cm]{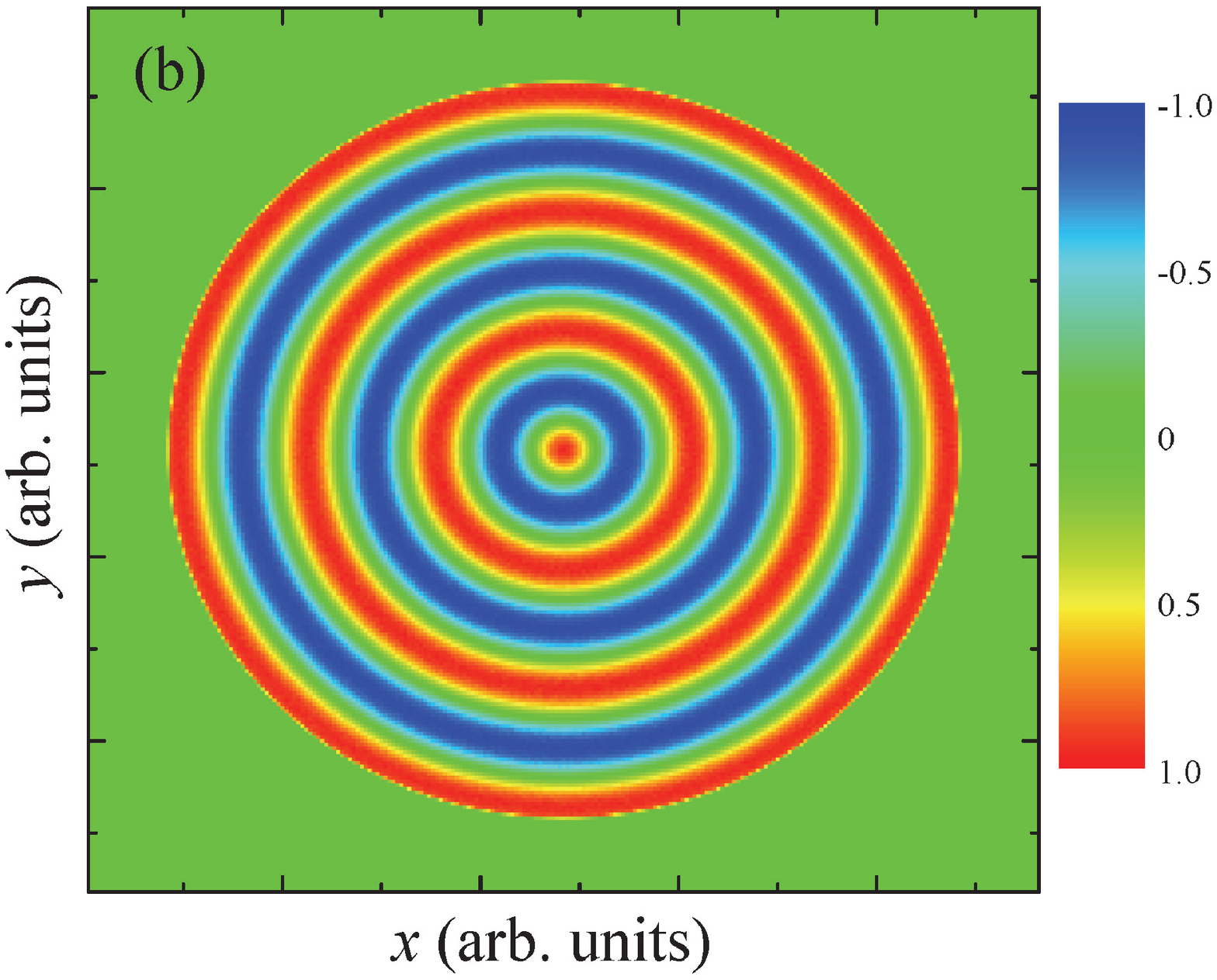}
\caption{(Color online) (a) Schematic of initial spin polarization
distribution in a radial spin helix. (b) Initial distribution of
$z$-component of spin polarization ($S_z(r,0)/S_0$) in the radial
spin helix of finite radius used in our Monte Carlo simulations.}
\label{fig1}
\end{figure}

In this paper, we consider spin relaxation of a {\it radial spin
helix} in a 2D electron system with Rashba spin-orbit (SO)
interaction. This structure is interesting because it provides, to
the best of our knowledge, the longest spin relaxation time of 2D
electrons subjected to Rashba SO interaction. Such a property is
related to spin polarization in the vicinity of the special point
of radial spin helix $r=0$. Physically, at short times, when the
period of radial spin helix is equal to the period of spin
precession, an electron diffusing from any direction to a point in
the vicinity of $r=0$ has the same direction of spin polarization
as the initial spin polarization at this point. Thus,
D'yakonov-Perel' spin relaxation becomes inefficient at short
times for spin polarization in the vicinity of $r=0$.

In the radial spin helix, the initial distribution of spin
polarization has a cylindrical (axial) symmetry and is given by
\begin{eqnarray}
S_r(r,t=0)=-S_0\sin(kr), \label{Sr} \\ S_\varphi(r,t=0)=0,
\\ S_z(r,t=0)=S_0\cos(kr) \label{Sz},
\end{eqnarray}
where $k$ is the wave vector, and $S_0$ is the initial amplitude
of spin polarization. In Fig. \ref{fig1}(a) we show schematically
spin polarization distribution in the radial direction. The
overall distribution of $z$-component of the spin polarization in
radial spin helix of finite radius is shown in Fig. \ref{fig1}(b).
Moreover, in this paper we will often refer to a plain spin helix
suggested in Ref. \onlinecite{Pershin05a}. The initial
distribution of spin polarization components in the plain spin
helix is
\begin{eqnarray}
S_x(x,t=0)=-S_0\sin(kx), \label{Splnx} \\ S_y(x,t=0)=0,
\\ S_z(x,t=0)=S_0\cos(kx) \label{Splnz}.
\end{eqnarray}

This paper is organized as follows. In Sec. \ref{sectionDD}, a
drift-diffusion equation approach is used to study relaxation
dynamics of the radial spin helix. We obtain general expressions
for spin polarization as a function of time and study its short
and long time behavior. Our analytical studies are supported by
Monte Carlo simulations presented in Sec. \ref{sectionMC}. Our
main results and conclusions are summarized in Sec.
\ref{sectionConcl}. Moreover, in several Appendices following the
main text, we provide additional calculation details.
Specifically, electron spin rotations induced by Rashba SO
interaction are considered in the Appendix \ref{app:rot}, spin
drift-diffusion equations are derived in the Appendix
\ref{app:ddscheme}, a general analytical solution of spin
drift-diffusion equation in the axially symmetric case is
presented in the Appendix \ref{app:sol}, and relaxation of plain
spin helix is discussed in the Appendix \ref{app:plain}.

\section{ Drift-diffusion description of radial spin helix} \label{sectionDD}

In this section we consider  the relaxation dynamics of the radial
spin helix analytically. The initial spin polarization in the
radial spin helix is of cylindrical symmetry and described by Eqs.
(\ref{Sr}-\ref{Sz}). Intuitively, a special point in the radial
spin helix is $r=0$, because the electrons motion through this
point along straight trajectories in {\it all} directions should
not lead to spin relaxation at short times for a specific value of
the wave vector $k$. Correspondingly, the spin lifetime of
electrons located in a region within $r=0$ should be longer than
the spin lifetime of homogeneous spin polarization and of plain
spin helix. This effect is in the focus of our investigation.

Let us consider a two-dimensional electrons confined in a quantum
well or heterostructure with Rashba-type spin-orbit interaction
\cite{Bychkov84a}. The standard Hamiltonian with the Rashba term
is given by
\begin{equation}
\hat H=\frac{\mathbf{\hat p}^2}{2m}+\alpha\left(\mathbf{\hat\sigma}
\times\mathbf{\hat p}\right)\cdot\mathbf{z},
\label{ham}
\end{equation}
where $\mathbf{\hat{p}}=(\hat p_x,\hat p_y)$ is the 2D electron
momentum operator, $m$ is the effective electron's mass,
$\mathbf{\hat\sigma}$ is the Pauli-matrix vector, $\alpha$ is the
spin-orbit coupling constant and $\mathbf{z}$ is a unit vector
perpendicular to the confinement plane.

It is not difficult to show that in the case of Hamiltonian
(\ref{ham}) the quantum mechanical evolution of a spin of an
electron with a momentum $\mathbf{p}$ can be reduced to a spin
rotation with the angular velocity $\Omega=2\alpha p/\hbar$ about
the axis determined by the unit vector
$\mathbf{n}=\mathbf{p}\times \mathbf{z}/p$ (see Appendix
\ref{app:rot}). In this way, the spin-orbit coupling constant
$\alpha$ enters into equations through the parameter $\eta=2\alpha
m \hbar^{-1}$, which gives the spin precession angle per unit
length.

Besides this evolution, 2D electrons experience different bulk
scattering events such as, for example, due to phonons or
impurities. These scatterings randomize the electron trajectories.
Correspondingly, the direction of spin rotation becomes
fluctuating what causes average spin relaxation (dephasing). This
is the famous D'yakonov-Perel' spin relaxation
mechanism.~\cite{Dyakonov72a,Dyakonov86a} The time scale of the
bulk scattering events can then be characterized by a single rate
parameter, the momentum relaxation time $\tau$. It is connected to
the mean free path by $\ell=v \tau$, where $v=p/m$ is the mean
electron velocity. To take into account  these scatterings we use
a model of diffusive spin transport, which in the limit of
small $k\ell\ll 1$, yields the spin
drift-diffusion equations (\ref{SxEq}-\ref{SzEq}) (Appendix
\ref{app:ddscheme} provides derivation details).

Let us consider dynamics of a radial spin helix relaxation. We
assume that such a structure is created at the initial moment of
time with the spin polarization components given by Eqs.
(\ref{Sr}-\ref{Sz}). The exact solution of the radial spin
drift-diffusion equations (\ref{SrEq}-\ref{SzrEq}) with initial
conditions (\ref{Sr}-\ref{Sz}) and constants $\gamma$ and $C$
from Eqs. (\ref{constants}) can be written as (see Appendix
\ref{app:sol} for more details)
\begin{eqnarray}
\frac{S_r(r,t)}{S_0}=-\frac{d}{dk}\int_0^k\frac{dsJ_1(sr)}{\sqrt{k^{2}-s^{2}}}\left[ k\cosh\left(\sqrt{\eta^2+16s^2}\frac{\eta Dt}{2}\right)\right.\nonumber\\ +
\left.\left(k\eta+4s^2 \right)
\frac{\sinh\left(\sqrt{\eta^2+16s^2}\frac{\eta Dt}{2}\right)}
{\sqrt{\eta^2+16s^2}}
 \right]e^{-(s^2+3\eta^2/2)Dt},~~~~~ \label{Srf}
\\
\frac{S_z(r,t)}{S_0}=\frac{d}{dk}\int_0^k\frac{dssJ_0(sr)}{\sqrt{k^{2}-s^{2}}}\left[ \cosh\left(\sqrt{\eta^2+16s^2}\frac{\eta Dt}{2}\right)\right. \nonumber\\
\left.+\left(4k-\eta\right)
\frac{\sinh\left(\sqrt{\eta^2+16s^2}\frac{\eta Dt}{2}\right)}
{\sqrt{\eta^2+16s^2}}
 \right]e^{-(s^2+3\eta^2/2)Dt},~~~~~~~\label{Szf}
\end{eqnarray}
where $J_1(r)$ and $J_0(r)$ are the Bessel functions of the first
and zeroth order correspondingly and $D$ is the diffusion
constant.

Eqs. (\ref{Srf}-\ref{Szf}) define completely  the radial spin
helix at any point $r$ and at any moment of time $t$.  The time
dependence of spin polarization at the center of helix is of
particular interest because the spin relaxation at this point is
the slowest. The radial component of spin polarization $S_r$ in
the vicinity of $r=0$ is close to zero (it follows from symmetry
considerations or directly from Eq. (\ref{Srf})). Therefore,
below, we derive asymptotic expressions at short and long times
for $S_z$ only. At short times $D\eta^2 t\ll 1$, an
 expansion of the RHS of Eq. (\ref{Szf}) in $t$ and its integration over $s$ at
 $r=0$ results in
 \begin{eqnarray}
\frac{S_{z}(0,t)}{S_{0}}=1-2(k-\eta)^2 Dt+\frac{2}{3}(k-\eta)(2k^3-6k^2\eta~~~~~~~~~~~~~\nonumber\\
 +6k\eta^2-3\eta^3)D^2t^2-\left(\frac{8}{15}k^6-\frac{16}{5}k^{5}\eta+8k^4\eta^2-\frac{104}{9}k^3\eta^3\right.~~~\nonumber\\
\left.  +\frac{32}{3}k^2\eta^4-\frac{14}{3}k\eta^5+\frac{4}{3}\eta^6\right)D^3t^3+O(t^4).~~~~~~~~~~~~
 \label{Sz_Small_t}
 \end{eqnarray}
It follows from Eq. (\ref{Sz_Small_t}) that when the radial spin
helix period is equal to spin precession length (this happens when $k=\eta$)
the decay of $S_z$ at the center of helix starts with a cubic term
in $t$
\begin{equation}
 \frac{S_{z}(0,t)}{S_{0}}=1-\frac{10}{9}(D\eta^2t)^3+O(t^4),~
 \text{for } k=\eta, D\eta^2t\ll 1.
 \label{Sz_Small_t2}
\end{equation}
This means that the spin relaxation at the center of the radial
spin helix at short times is significantly suppressed (for this
special wave number, $k=\eta$) and characterized by a rather long
initial interval of non-exponential behaviour.

The asymptotic behaviour of $S_r$ and $S_z$ at long times,
$D\eta^2 t\gg 1$, can be determined by taking into account the
dominant contribution to the integrals in Eqs. (\ref{Srf}) and
(\ref{Szf}). This contribution comes from the vicinity of a point
$s\in[0,k]$ corresponding to the maximum of the $-\lambda_{-}(s)$
in the integration interval (see Eq. (\ref{lambda1}) and Fig.
\ref{fig7}). We should consider three cases.  In the first case,
when $0<k< k_m=\sqrt{15}\eta/4$, the main contribution to the
integrals in Eqs. (\ref{Srf}) and  (\ref{Szf}) comes from the
vicinity of point $s=k$ at the right end of the integration
interval. In the second case, when $k=k_m$, we should keep in mind
that $k=k_m$ is a stationary point of $\lambda_{-}(s)$ entering
the exponent. Therefore, in this case, the asymptotic behaviour
differs from the case $k<k_m$ by a pre-exponential factor. In the
third case, when $k>k_m$, the main contribution to the integrals
in Eqs. (\ref{Srf}) and  (\ref{Szf}) arises from the vicinity of
the inner stationary point $s=k_{m}$  of $\lambda_{-}(s)$.

The asymptotic behaviour of Laplace-type integrals is obtained
using standard for this purpose technics. From Eq. (\ref{Szf}) we
get
\begin{eqnarray}
 \frac{S_{z}(r,t)}{S_{0}}=\left( 1+\frac{4k-\eta}{\sqrt{\eta^2+16k^{2}}} \right)J_0(kr)\sqrt{-\frac{\pi k}{8}\frac{d\lambda_{-}(k)}{dk}t}\nonumber\\
\times e^{-\lambda_{-}(k)t},~~
 \text{for }~ 0<k<\frac{\sqrt{15}}{4}\eta,~ -k\frac{d\lambda_{-}(k)}{dk}t
 \gg 1,~~~
 \label{Sz_Big_t1}
\end{eqnarray}
\begin{eqnarray}
 \frac{S_{z}(r,t)}{S_{0}}=\frac{\sqrt{15}(\sqrt{15}+3)}{32\sqrt{2}}\Gamma\left( \frac{3}{4}\right)J_0(k_mr)(D\eta^2 t)^{\frac{1}{4}}\nonumber\\
\times e^{-\lambda_{-}(k_{m})t},~~
 \text{for }~ k=k_{m}=\frac{\sqrt{15}}{4}\eta,~ D\eta^2 t\gg 1,~
 \label{Sz_Big_t_km}
\end{eqnarray}
\begin{eqnarray}
 \frac{S_{z}(r,t)}{S_{0}}=-\frac{3}{32}\frac{(4k+5\eta)\eta}{(k^2-k_m^2)^{3/2}}J_0(k_mr)
 \sqrt{\frac{\pi }{Dt}}e^{-\lambda_{-}(k_{m})t},\nonumber\\
 \text{for }~ k>\frac{\sqrt{15}}{4}\eta,~ (k-k_m)^2Dt
 \gg 1,~D\eta^2 t \gg 1.~~~~~
 \label{Sz_Big_t2}
\end{eqnarray}

From Eq. (\ref{Sz_Big_t1}) we see that if the wave vector $k$
satisfies the inequality $0<k<k_m$, then, at long times, the spin
polarization decay is mainly exponential and characterized by a
relaxation time $\tau(k)=(\lambda_{-}(k))^{-1}$. This relaxation
time increases with $k$. At the same time (accordingly to Eq.
(\ref{Sz_Small_t})), the spin polarization decay at short times
also decreases with $k$. Therefore, the spin life time of the
radial spin helix increases with increase of $k\in(0, k_m)$.

\begin{figure}[tb]
\centering
\includegraphics[width=7.5cm]{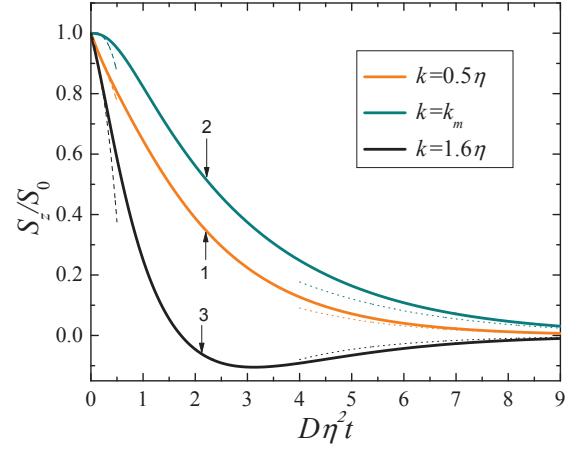}
\caption{(Color online) Time-dependence of $S_z(0,t)$ in the
radial spin helix at several values of $k$.} \label{fig2}
\end{figure}

At $k=k_m$, as it follows from Eqs.
(\ref{Sz_Big_t_km}-\ref{Sz_Big_t2}), the relaxation time reaches
its maximum value
$\tau_m=(\lambda_{-}(k_m))^{-1}=(7D\eta^2/16)^{-1}$. Moreover,
since $k_m\approx0.97\eta$ is very close to $\eta$, the conditions
for short time suppression of spin relaxation are almost optimal
at this value of $k$. Therefore, when $k=k_m$, the spin relaxation
is significantly suppressed at both short and long times.

When $k>k_m$, the asymptotic relaxation time is the same as when
$k=k_m$ (see Eqs. (\ref{Sz_Big_t_km},\ref{Sz_Big_t2})). However,
even when $k$ is close to $k_m$, the ratio of the absolute value
of RHS of Eq. (\ref{Sz_Big_t2}) to those of Eq.
(\ref{Sz_Big_t_km}) is small (of the order of
$[(k-k_m)^2Dt]^{-3/4}$). In addition, in the asymptotic  formula
(\ref{Sz_Big_t2}), the pre-exponential factor changes its sign.
Therefore, the spin polarization $S_z$ must turn to zero at some
moment of time, before it reaches the asymptotic behaviour given
by Eq. (\ref{Sz_Big_t2}). Eq. (\ref{Sz_Small_t}) at $k>\eta\approx
k_m$ also predicts a relaxation increase with $k$ when $k>k_m$.

\begin{figure*}[t]
 \begin{center}
  \centerline{
    \mbox{(a)}
    \mbox{\includegraphics[width=7.5cm]{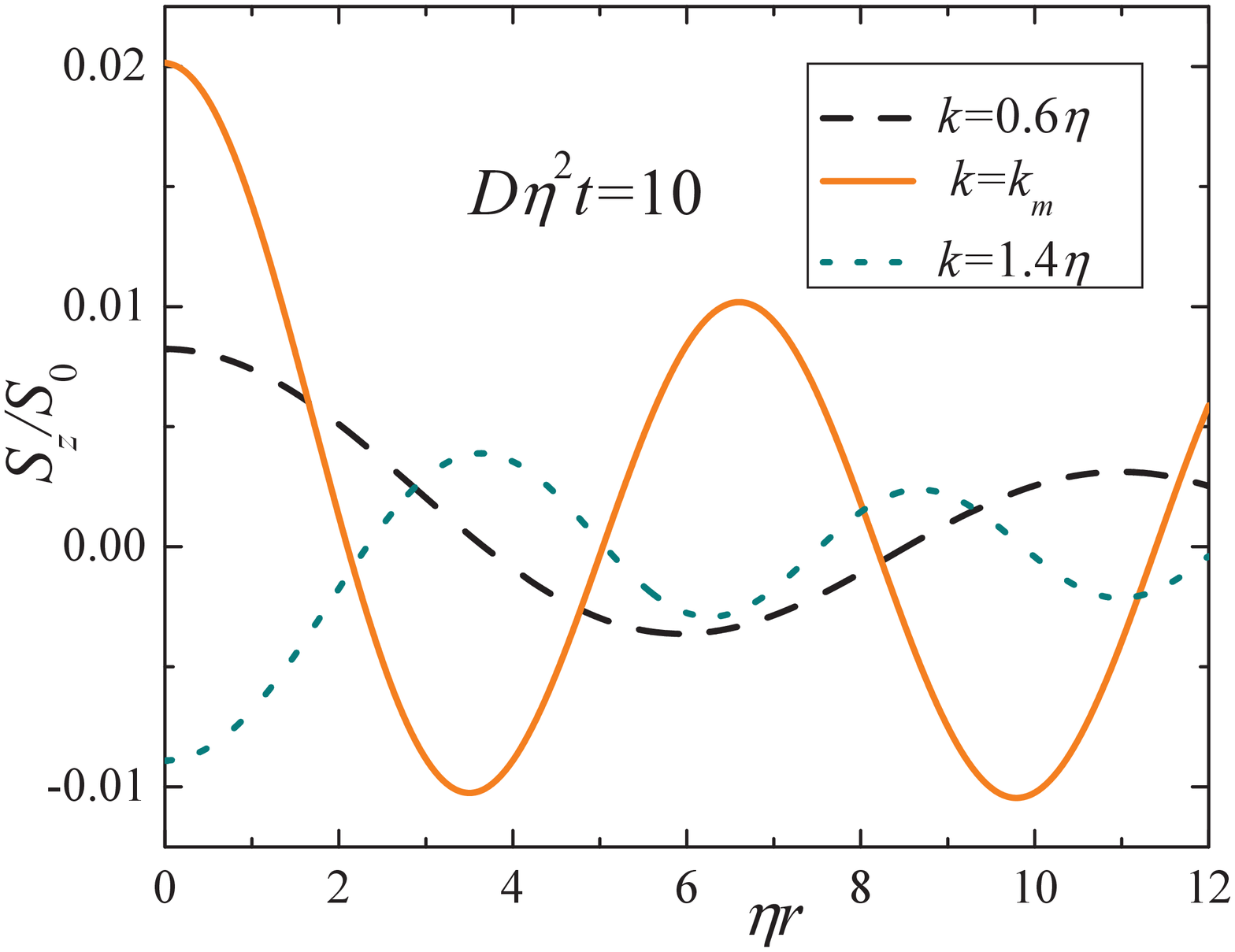}}
    \mbox{(b)}
    \mbox{\includegraphics[width=7.5cm]{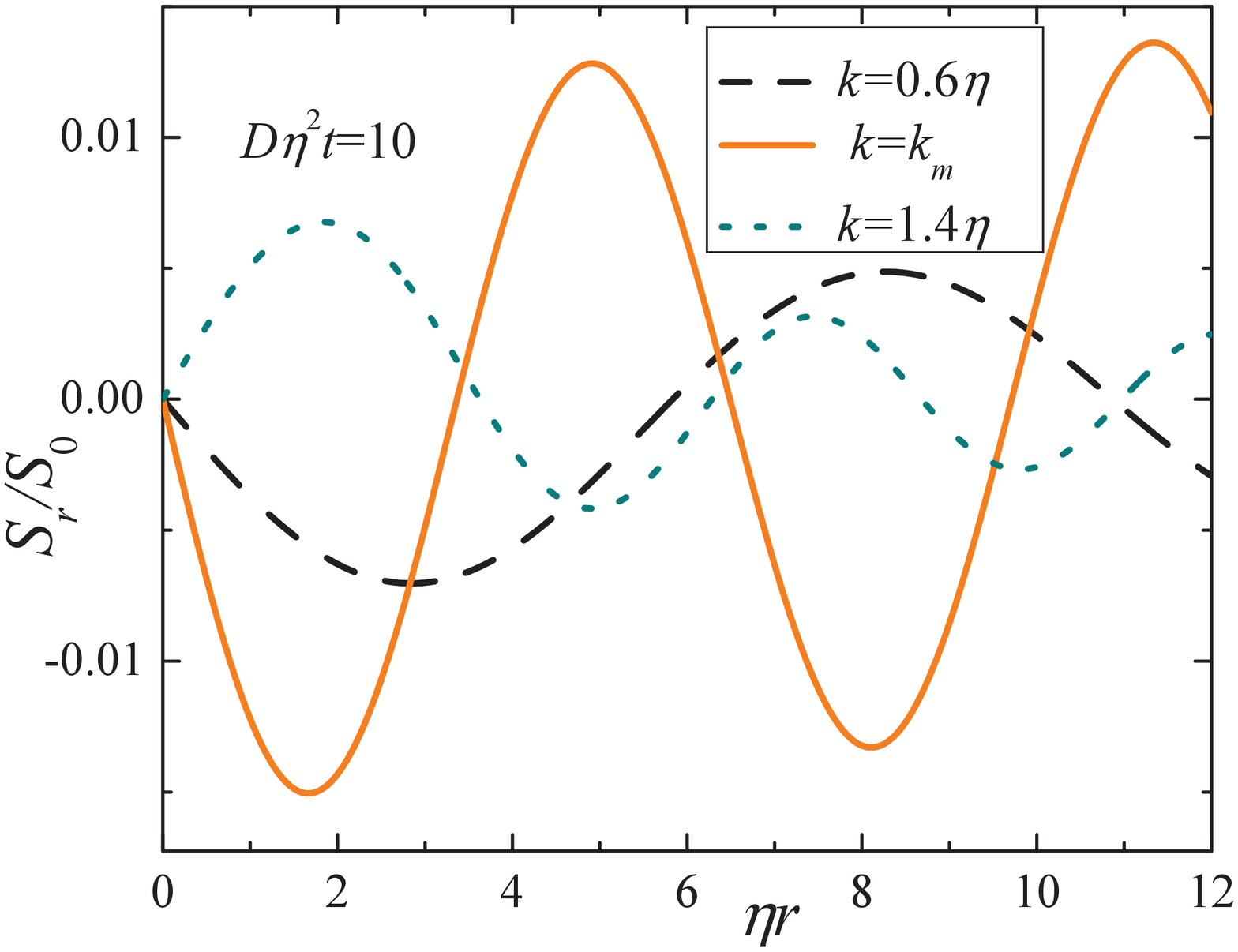}}
  }
\caption{ Spatial dependence of $S_z$ (a) and $S_r$ (b) in the
radial spin helix at the indicated moment of time.
}\label{fig_spatial}
\end{center}
\end{figure*}

Thus we conclude that the longest spin relaxation time for the
spin polarization at the center of the radial spin helix occurrs
at the wave vector
\begin{equation}
k=k_m=\sqrt{15}\eta/4 \label{k_mm}
\end{equation}
and is given by
\begin{equation}
\tau_m=(7D\eta^2/16)^{-1}. \label{tau_m}
\end{equation}
In addition, at this value of $k$, the dynamics of spin
polarization in the vicinity of $r=0$ is non-exponential at short
times, when the spin polarization remains almost constant. These
are the main results of our calculations.

The relaxation of initially homogeneous spin polarization (when
$k=0$) can be obtained from Eq. (\ref{Szf}) in the limit
$k\rightarrow 0$. In this limiting case the factor before the
exponential function $e^{-\lambda_{-}(s)t}$ in Eq. (\ref{Szf})
turns to zero in $k=0$ limit. Therefore, at $k=0$, the spin
relaxation is determined by the exponential function
$e^{-\lambda_{+}(0)t}$, where, accordingly to Eq. (\ref{lambda1}),
$\lambda_{+}(0)=2D\eta^2$.

In fact, the exact time dependence of $S_z$ at $k=0$ coincides
with its asymptotic behaviour. It can be clearly seen from both
Eqs. (\ref{Szf}) and (\ref{SzEq}) that
\begin{equation}
S_{z}(t)=S_0e^{-\lambda_{+}(0)t}=S_0e^{-2D\eta^2t},~ \mbox{for
}k=0. \label{homog}
\end{equation}
Accordingly to  Eq. (\ref{homog}), the initially homogeneous
spin polarization, directed along $z$-axis, decays exponentially
with a time constant
 $\tau^{(h)}=(2D\eta^2)^{-1}$.
We also note that the applicability limits of the long times
asymptotic expressions listed in Eqs.
(\ref{Sz_Big_t1}-\ref{Sz_Big_t2}) do not allow calculations of
$S_z$ at $k=0$ or $k=k_m$ as a limiting case of Eq.
(\ref{Sz_Big_t1}).

In Fig. \ref{fig2} we show the time dependence of the spin
polarization component $S_z(r=0,t)$ calculated at several wave
vectors ($k=0.5\eta$ (solid line 1), $k=k_m=\sqrt{15}\eta/4$
(solid line 2) and $k=1.6\eta$ (solid line 3)) using Eq.
(\ref{Szf}). The dashed lines 1, 2 and 3 represent asymptotic
behavior at short times (given by Eq. (\ref{Sz_Small_t})), and dotted  lines 1,
2, 3 show asymptotic behavior
at long times (given by Eqs. (\ref{Sz_Big_t1}-\ref{Sz_Big_t2})) for the same values of the wave vectors $k$. These curves
reveal main features discussed above.

The spatial dependences of $z$- and $r$-components of spin
polarization, calculated from Eqs. (\ref{Szf}) and (\ref{Srf}),
are presented in Fig. \ref{fig_spatial} for a particular moment of
time $D\eta^2t=10$ and wave vectors $k=0.6\eta$ (dashed lines),
$k=k_m$ (solid lines), and $k=1.4\eta$ (dotted lines). These plots
clearly demonstrate that the maxima of $S_z$ and $S_r$ are reached at the
wave vector $k=k_m$.

\begin{figure}[t]
\centering
\includegraphics[width=7.5cm]{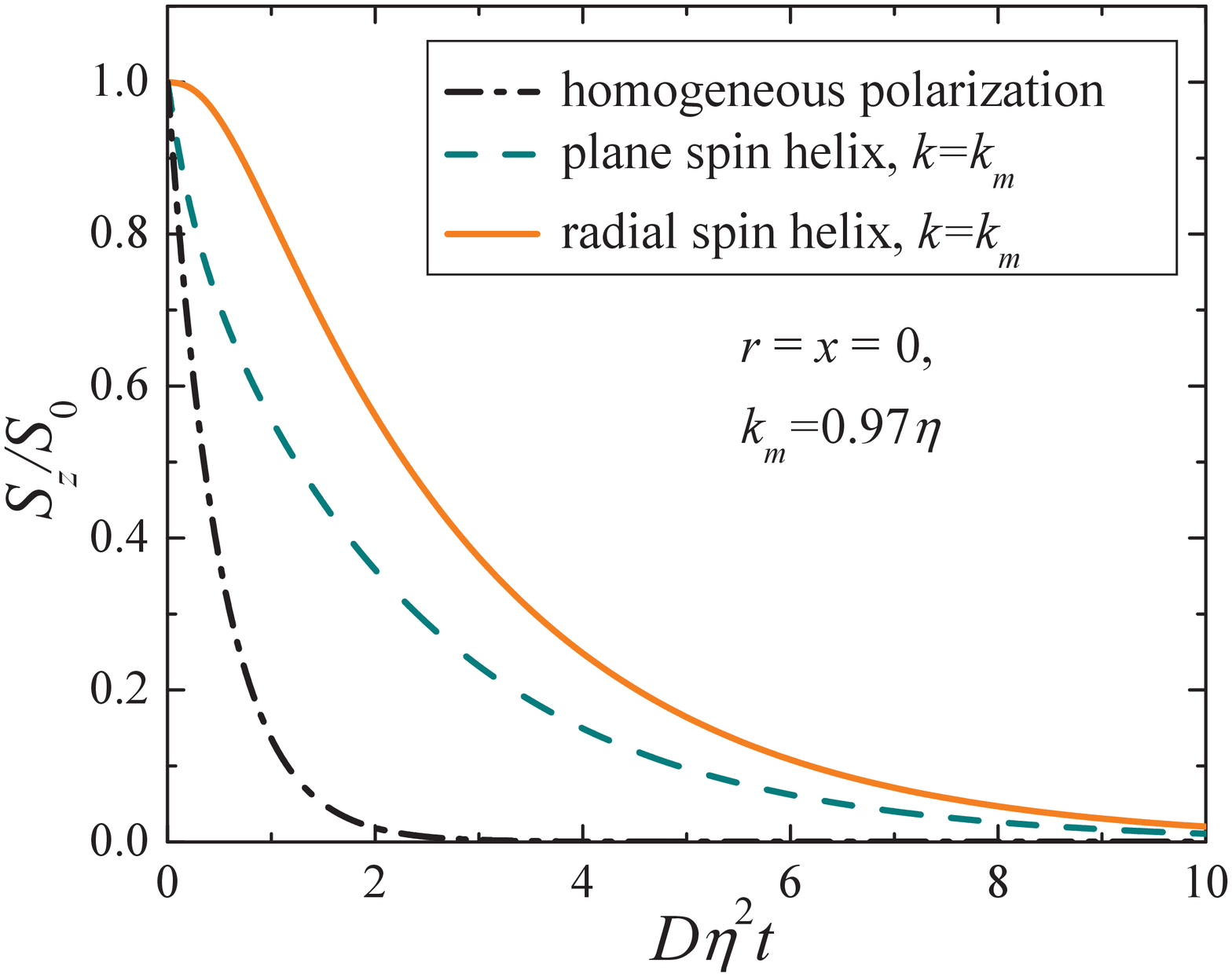}
\caption{(Color online) Time dependence of $S_z(r=0,t)$ for three
different initial spin configurations: homogeneous polarization,
plain spin helix and radial spin helix. The radial spin helix is
characterized by the largest magnitude of spin polarization at any
moment of time.} \label{fig3}
\end{figure}

Fig. \ref{fig3} depicts time dependence  of $S_z(r=0,t)$
for homogeneous spin polarization (calculated from Eq.
(\ref{homog})), plain spin helix (calculated from Eq.
(\ref{SzPlane})) and radial spin helix. This plot
demonstrates that the spin polarization at $r = 0$ in the radial
spin helix lives longer that those in the case of homogeneous
spin polarization and plain spin helix. It is interesting and
important that at short times this curve stays almost flat as
expected. At longer times, when the initial non-exponential
relaxation region ends, the relaxation of spin polarization in the
radial spin helix occurs with the same time constant as those of
the plain spin helix (it can be shown analytically from Eqs.
(\ref{SxPlane}-\ref{SzPlane}) that the longest spin relaxation
time of the plain spin helix is given by Eq. (\ref{tau_m}) and
occurs at $k=k_m$ given by Eq. (\ref{k_mm})). Therefore, in both
cases, the increase of the exponential relaxation time relative to the
homogeneous spin polarization is equal to $32/7$.

\section{Monte Carlo simulations} \label{sectionMC}

In order to obtain an additional insight on spin relaxation of the
radial spin helix, we perform Monte Carlo simulations employing an
approach described in Refs. \onlinecite{Kiselev00a,Saikin05a}.
This Monte Carlo simulation method uses a semiclassical
description of electron space motion and quantum-mechanical
description of spin dynamics (the later is based on the Hamiltonian (\ref{ham})).
All specific details of the Monte Carlo simulations
program can be found in the references cited above and will not be
repeated here. To some extent, Monte Carlo simulations program
numerically solves Eqs. (\ref{SxEq}-\ref{SzEq}) taking into
account relations (\ref{S_Rotation}) and (\ref{Omega_n}).

 \begin{figure}[t]
 \centering
 \includegraphics[width=7.5cm]{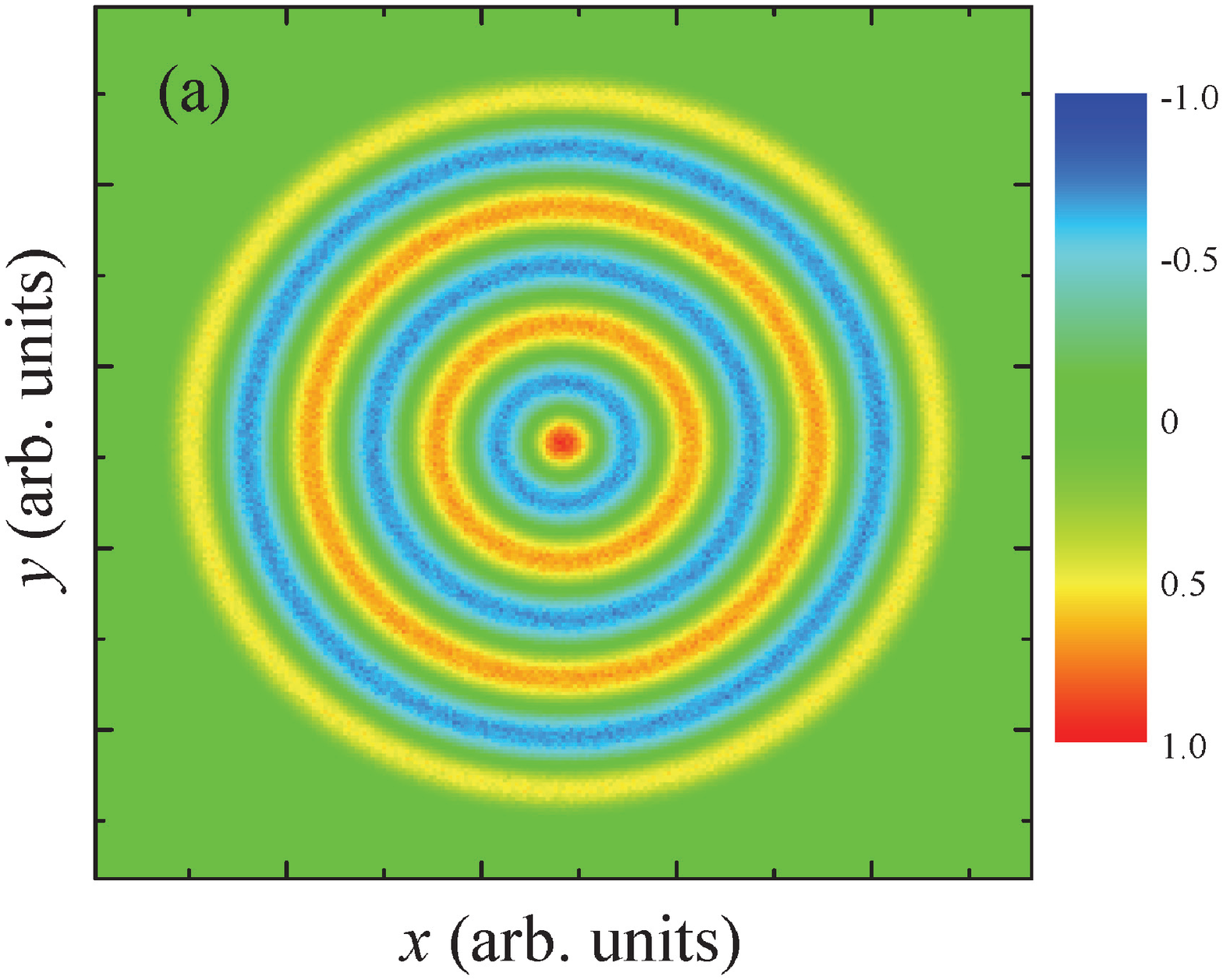}
 \includegraphics[width=7.5cm]{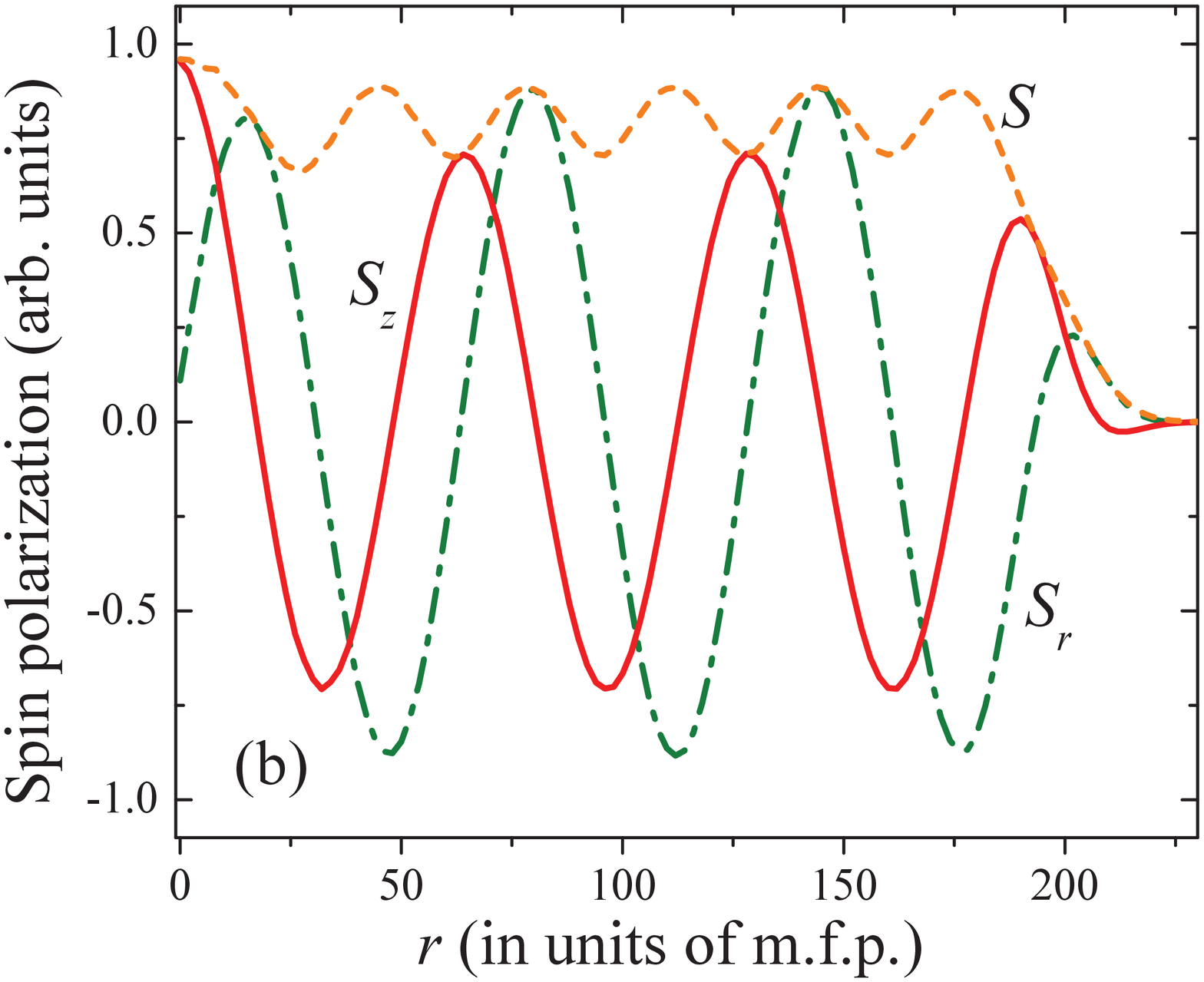}
 \caption{(Color online) Distributions of $z$ component (a) and all
 components (b) of spin polarization ($S_\varphi=0$,
 $S=\left(S_r^2+S_z^2 \right)^{1/2}$) at $t=100 \tau$ in the radial
 spin helix. The red (dark) area in the
 center of radial spin helix in (a) and the maximum of $S$ and $S_z$
 in the vicinity of $r=0$ in
 (b) demonstrate a longer spin lifetime of electrons located in
this region. This plot was obtained at $\eta \ell=0.1$ and the radial
spin helix period $a=64.77\ell$ (this value of $a$ corresponds to
$k=k_m$). M.f.p. (mean free path) stands for $\ell$.} \label{fig5}
\end{figure}

All numerical results related to the radial spin helix were
obtained using an ensemble of $10^8$ electrons initially
homogeneously distributed within a circle of a sufficiently large
radius $R=200\ell$ to insure that the influence of boundary
effects on spin polarization in a region in the vicinity of $r=0$ is
negligible. The initial configuration of $z$-component of spin
polarization of these electrons is presented in Fig.
\ref{fig1}(b). Our Monte Carlo simulations results are
in an excellent agreement with the theory of radial spin helix relaxation
presented above.

Fig. \ref{fig5}(a) demonstrates the spatial dependence of $S_z$ in
the radial spin helix at $t=100\tau$ and
Fig. \ref{fig5}(b) depicts the radial dependence of the spin
polarization components at the same moment of time. In particular,
it can be clearly seen that the decay of $S_z$ in the vicinity of
$r=0$ is slower than in other regions. The total spin polarization
$S$ shows oscillations that are related to a well-known feature of
D'yakonov-Perel' relaxation: the spin relaxation time of the
perpendicular to plane spin polarization is shorter than the spin
relaxation time of the in-plane spin polarization. Similar
oscillation of spin polarization amplitude were previously
found in the relaxation dynamics of the plain spin
helix.~\cite{Pershin05a}

In addition to the radial spin helix relaxation, we simulated the
relaxation of homogeneous spin polarization and relaxation of
plain spin helix. Monte Carlo simulations reveal that at long
times the time dependence of spin polarization in all spin
configurations (homogeneous spin polarization, plain spin
helix  and radial spin helix) exhibits an exponential
decay.

\begin{figure}[t]
 \centering
 \includegraphics[width=7.5cm]{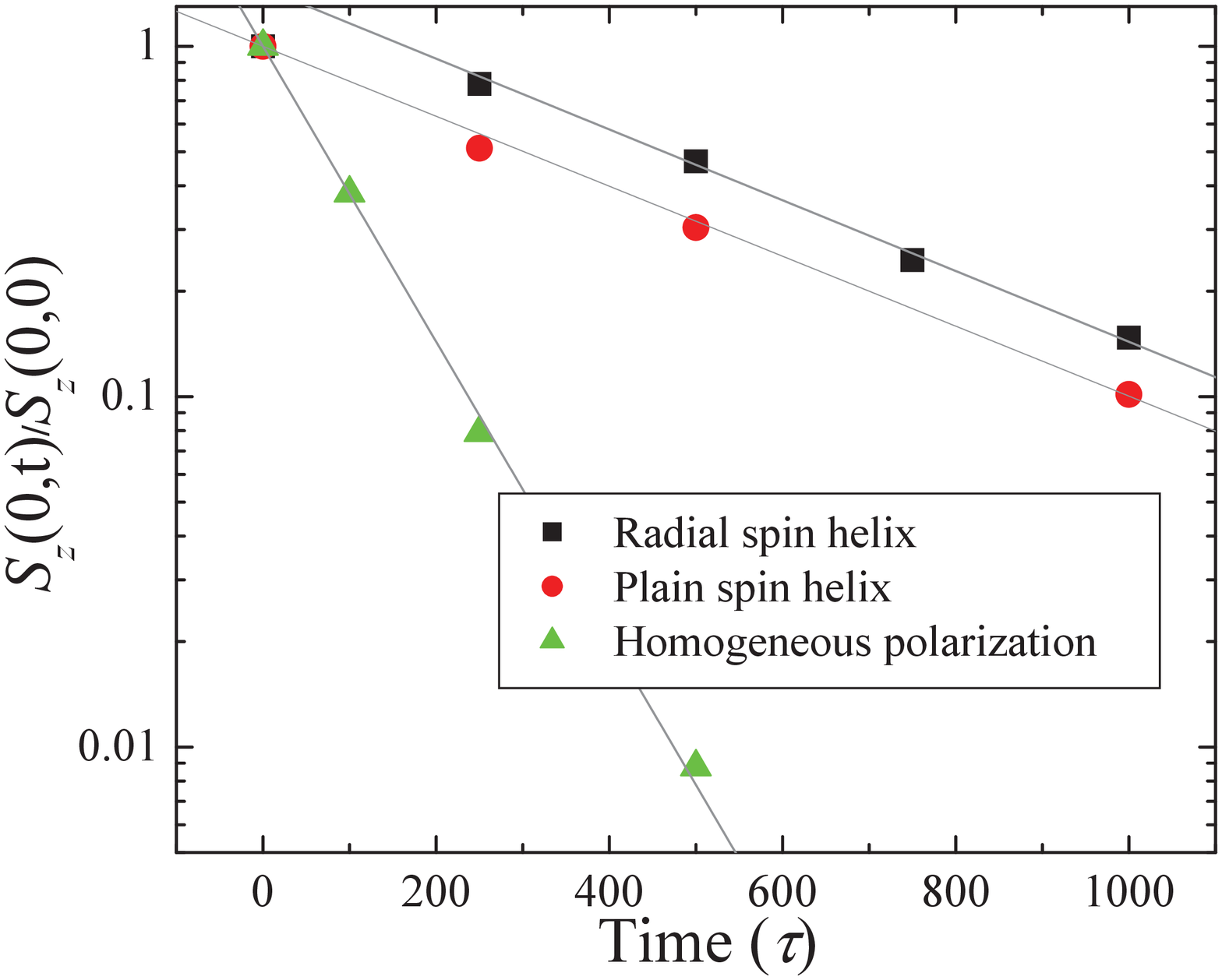}
 \caption{(Color online) Spin polarization as a function of time for different
 initial spin polarization configurations. This
 plot was obtained using the parameters values $\eta \ell=0.1$ and
 $a=64.77 \ell$ (the period of helices). The straight lines are fitting curves selected as
 $\textnormal{exp}(-(t-t_0)/t_1)$.
  The parameters of the fitting curves are $t_0=0$, $t_1=103\tau$ for
 homogeneous spin polarization, $t_0=0\tau$, $t_1=435\tau$ for
 plain  spin helix, and $t_0=165\tau$, $t_1=430\tau$ for radial spin helix.}
 \label{fig6}
 \end{figure}

In Fig. \ref{fig6} we compare the time dependencies of spin
polarization ($S(0,t)/S_0$) for three different initial
polarization configurations: the homogeneous spin polarization
(initial spin polarization is selected as $S_x=S_y=0$, $S_z=S_0$),
plain spin helix and radial spin helix. The selected period of
radial and plane spin helices ($k=k_m$) corresponds to the longest
spin lifetime of these structures at a fixed $\eta \ell=0.1$. It
follows from Fig. \ref{fig6} that the spin polarization in the
radial spin helix is the most robust against the relaxation. We
emphasize that the long-time behavior of all curves can be
perfectly fitted by an exponential law given in the caption of
Fig. \ref{fig6}. The numerically obtained increase in the spin
lifetime of the radial spin helix $435/103\sim 4.2$ is very close
to the theoretically predicted value $32/7 \sim 4.6$. We also note
that a slightly longer spin relaxation time increase ($\sim 6$)
reported in Ref. \onlinecite{Pershin05a} for the plain spin helix
can be related to a large value of $\eta \ell=0.3$ that is beyond the
linear spin drift-diffusion theory.

\section{Conclusion} \label{sectionConcl}

The dynamics of spin relaxation of radial spin helix was
investigated using spin drift-diffusion equations and Monte Carlo
simulations. Starting with a clear model of diffusive spin transport, we
derived spin drift-diffusion equations for electron spin
polarization in 2D semiconductor structures with Rashba spin-orbit
coupling. The general solution of these equations for the axial
symmetric case was found. Based on this solution, we studied the evolution of
the suggested long-lived spin structure - the radial spin helix.
It was shown that the relaxation of spin polarization in the vicinity of $r=0$ in this structure
demonstrates an unusual long initial non-exponential relaxation behavior followed by an exponential decay.
The optimal value of the radial spin helix wave vector was found and
corresponding exponential relaxation time was calculated.
Qualitatively, the initial non-exponential decay feature can be explained by
existence of an infinite set of dephasing-free trajectories
propagating through the point $r=0$.

In order to additionally check our analytical results, we
also performed Monte Carlo simulations of the dynamics of radial
spin helix relaxation using the same Monte Carlo simulation technique as those described in
Refs. \onlinecite{Kiselev00a,Saikin05a}. Fig. \ref{fig5} shows a
representative result of our simulation in which it is clearly
demonstrated that the polarization decay at $r = 0$ is slowest.
Our analytical and Monte Carlo simulations results are in a perfect
agreement.

To conclude, the radial spin helix is a new structure
exhibiting an unusual spin relaxation dynamics and relatively long lifetime.
Its property of slow relaxation dynamics at short times is very interesting.
Experimentally, the radial spin helix
can be created by spin injection from a point electrode located at
the center of a second ring-shape electrode or possibly by a
modified spin gratings technique \cite{Cameron96a}.

\appendix

\section{Rotation of a single spin subjected to Rashba SO
interaction} \label{app:rot}

Let us consider an electron with a momentum
$\mathbf{p}=(p_x,p_y)$. The Hamiltonian (\ref{ham}) can be
rewritten as
\begin{equation} \hat H=\frac{\mathbf{p}^2}{2m}+\alpha
p\mathbf{n\cdot}\mathbf{\hat\sigma}, \label{Ham_spin}
\end{equation}
where $\mathbf{n}=\mathbf{p}\times \mathbf{z}/p$ is the unit
vector.

Since eigenvalues spin projection operator $\mathbf{\hat s\cdot
n=\hat\sigma\cdot n}/2$ on any direction are equal to $\pm 1/2$,
the energy levels of (\ref{Ham_spin}) are given by
\begin{equation}
E_{\pm}=\frac{p^2}{2m}\pm \alpha p, \label{Schrodinger}
\end{equation}
and it is clear that corresponding spin eigenfunctions are the
states with spin directed along and opposite to $\mathbf{n}$.

The evolution operator acting only on spin variables of the
electron with momentum $\mathbf{p}$ is equal to
\begin{equation}
\hat U(t)=e^{-i\hat H t/\hbar}= \exp\left\{\frac{-i\mathbf{p}^2
t}{2m\hbar}\right\} \exp\left\{-i\Omega
t\mathbf{n}\cdot\frac{\mathbf{\hat\sigma}}{2}\right\},~~~
\label{Smatrix}
\end{equation}
where $\Omega=2\alpha p/\hbar$.

We note that the evolution operator (\ref{Smatrix}) coincides with
the operator of finite rotation by an angle $\Omega t=2\alpha p
t/\hbar$ about $\mathbf{n}$-axis up to the phase factor
$e^{-ip^2t/(2m\hbar)}$. Since electron spin $s$ transforms under
rotations as the ordinary vector, we obtain
\begin{equation}
\mathbf{s}(t)=\cos(\Omega t)\mathbf{s}(0)+\sin(\Omega
t)\mathbf{n}\times\mathbf{s}(0)+ 
2\sin^2(\Omega t/2)\mathbf{n}\cdot\mathbf{s}(0)\mathbf{n},
 \label{S_Rotation}
\end{equation}
where
 \begin{equation}
 \Omega=2\alpha p/\hbar,~~\mathbf{n}=\mathbf{p}\times \mathbf{z}/p.
 \label{Omega_n}
\end{equation}

\section{Derivation of spin drift-diffusion equations}\label{app:ddscheme}

Let us consider a two-dimensional non-degenerate electron gas and
use a semiclassical approach to model the electron space motion
and quantum-mechanical approach based on the Hamiltonian
(\ref{ham}) to describe the electron spin dynamics. Moreover, we
assume the electrical neutrality and absence of an external
electromagnetic field. Within our approach, 2D electrons are
characterized by the momentum relaxation time $\tau$ and  the mean
free path $\ell$, so that the average velocity of electrons is
$v=\ell/\tau$. From elementary gas-kinetic considerations
\cite{Reif65a} we can write an equation for the change of electron
spin polarization $\Delta\mathbf{S}(x,y,t)$ in a region of
dimensions $2\ell \times 2\ell$ with the center at $(x,y)$ during
the time interval $\tau$:
\begin{eqnarray}
(2\ell)^2\Delta\mathbf{S}(x,y,t)= \frac{1}{4}v\tau
(2\ell)\left\{\mathbf{S}^\prime(x-2\ell,y,t)\right.
\label{S_flux}~~~~~\nonumber\\ +\mathbf{S}^\prime(x+2\ell,y,t)+
\mathbf{S}^\prime(x,y-2\ell,t)+\mathbf{S}^\prime(x,y+2\ell,t)~~~
\label{DDeqinit}
\\ \left.-4\mathbf{S}(x,y,t)\right\}.\nonumber
\end{eqnarray}

In the right hand side of Eq. (\ref{S_flux}), the first four terms
are the spin polarization fluxes into the region from four sides with
length $2\ell$, and the last term is the flux out of this region.
The prime symbols  in Eq. (\ref{S_flux}) denote a change of spin polarization
because of SO interaction-induced spin precession by the angle $2\Omega \tau=4\alpha m\ell/\hbar=2\eta\ell$ accordingly to Eqs. (\ref{S_Rotation},\ref{Omega_n}).
For example,
\begin{eqnarray}
\mathbf{S}^\prime(x-2\ell,y,t)=\cos(2\eta\ell)\mathbf{S}(x-2\ell,y,t)-
\sin(2\eta\ell)\mathbf{ y}~~\nonumber\\
\times\mathbf{S}(x-2\ell,y,t)+ 2\sin^2(\eta\ell)\mathbf{
y}\cdot\mathbf{S}(x-2\ell,y,t) \mathbf{ y},~~~ \label{S_prime}
\end{eqnarray}
where $\bf{y}$ is the unit vector along $y-$axis.

In order to obtain drift-diffusion equations for spin polarization, we
substitute expressions for $\mathbf{S}^\prime$ into Eq. (\ref{DDeqinit}), and expand
 trigonometrical functions up to quadric terms  with respect to small $2\eta \ell$
 and $\mathbf{S}^\prime$ terms up to  quadric terms with respect to $2\ell$. The resulting system of drift-diffusion equations
for spin polarization have a form
\begin{eqnarray}
\frac{\partial S_x}{\partial t}=D\Delta S_x+C\frac{\partial
S_z}{\partial x}-2\gamma S_x \label{SxEq}, \\ \frac{\partial
S_y}{\partial t}=D\Delta S_y+C\frac{\partial S_z}{\partial
y}-2\gamma S_y \label{SyEq}, \\
 \frac{\partial S_z}{\partial t}=D\Delta S_z-C\left(\frac{\partial S_x}{\partial
x}+\frac{\partial S_y}{\partial y}\right)-4\gamma S_z
\label{SzEq},
\end{eqnarray}
where
\begin{eqnarray}
C=2\eta D,~~\gamma=\frac{1}{2}\eta^2 D, \label{constants}
\end{eqnarray}
and
\begin{eqnarray}
D=\frac{\ell^2}{2\tau}. \label{Dh}
\end{eqnarray}
Here $D$ is the coefficient of diffusion, $C$ describes spin
rotations, and $\gamma$ is the coefficient describing spin
relaxation.

It is interesting to note that the same drift-diffusion equations
(\ref{SxEq}-\ref{constants}) can be obtained for the model of 2D
localized electrons on a lattice \cite{Pershin04b} in the hopping regime.  However, in this case,
the diffusion coefficient is equal to $D=\ell^2/(4\tau)$, where $\tau$ is the characteristic hopping time and
$\ell$ is the distance between lattice sites.

\section{Analytical solution of drift-diffusion equations in the
axially symmetric case} \label{app:sol}

In the axially symmetric case (assuming that $S_r=S_r(r,t)$,
$S_z=S_z(r,t)$ and $S_\varphi=0$) Eqs. (\ref{SxEq}-\ref{SzEq})
can be written as
\begin{eqnarray} \frac{\partial S_r}{\partial
t}=D\left[ \frac{\partial }{r\partial r}\left( r\frac{\partial
S_r}{\partial r} \right) -\frac{S_r}{r^2} \right]+C\frac{\partial
S_z}{\partial r}-2\gamma S_r ,~~\label{SrEq}
\\ \frac{\partial S_z}{\partial t}=D \frac{\partial }{r\partial
r}\left( r\frac{\partial S_z}{\partial r} \right) -C\frac{\partial
\left( r S_r \right)}{r\partial r}-4\gamma S_z . ~~\label{SzrEq}
 \end{eqnarray}

Let us find a general solution of Eqs. (\ref{SrEq}-\ref{SzrEq})
for the case of an infinite plane.  We search a specific solution
of the above Eqs. in the form
\begin{eqnarray}
  S_r(r,t)=A(s,t)J_1(sr)\label{SrSpecEq}, \\
  S_z(r,t)=B(s,t)J_0(sr)
 \label{SzrSpecEq},
 \end{eqnarray}
where $J_1(r)$ and $J_0(r)$ are the Bessel functions of the first
and zeroth order correspondingly. Substituting expressions
(\ref{SrSpecEq}-\ref{SzrSpecEq}) into Eqs.
(\ref{SrEq}-\ref{SzrEq}) we obtain a system  of  ordinary
differential equations for unknown functions $A(s,t)$ and $B(s,t)$
of positive parameter $s$ and time $t$
\begin{eqnarray}
  \frac{dA(s,t)}{dt}=-(Ds^2+2\gamma)A(s,t)-CsB(s,t)\label{AEq}, \\
   \frac{dB(s,t)}{dt}=-CsA(s,t)-(Ds^2+4\gamma)B(s,t)
 \label{BEq}.
 \end{eqnarray}
The general solution of  this system can be presented as
\begin{eqnarray}
 A(s,t)=C_+(s)sCe^{-\lambda_+(s)t}+C_-(s)sCe^{-\lambda_-(s)t}\label{ASolut}, \\
 B(s,t)=C_+(s)\left(\gamma+\sqrt{\gamma^2+C^2s^2}\right)e^{-\lambda_+(s)t}\label{BSolut}\\
   \nonumber
   +C_-(s)\left(\gamma-\sqrt{\gamma^2+C^2s^2}\right)e^{-\lambda_-(s)t},
 \end{eqnarray}
where we denote
\begin{equation}
 \lambda_{\pm}(s)=Ds^2
+3\gamma\pm\sqrt{\gamma^2+C^2s^2} \label{lambda},
\end{equation}
and $C_\pm(s)$ are  arbitrary functions of positive parameter $s$.
Using Eqs. (\ref{constants}), Eq. (\ref{lambda}) can be rewritten in a more simple
form
\begin{equation}
 \lambda_{\pm}(s)=\frac{1}{2}D(2s^2
+3\eta^2 \pm\eta\sqrt{\eta^2+16s^2}) \label{lambda1}.
\end{equation}

The special solutions (\ref{SrSpecEq}-\ref{SzrSpecEq}) (with
$A(s,t)$ and $B(s,t)$ given by Eqs. (\ref{ASolut}-\ref{lambda}))
of the radial drift-diffusion equations (\ref{SrEq}-\ref{SzrEq})
have the following simple meaning.  Accordingly to Eqs.
(\ref{SrSpecEq}-\ref{SzrSpecEq}), the spatial dependencies of the
radial and $z$-components of spin polarization are proportional to
the first and zeroth order Bessel function at any moment of time.
The parameter $s$ is similar to the wave vector $k$ for the plane
case. The amplitudes $A(s,t)$ and $B(s,t)$ determine the time
dependence of the radial and $z$-components of spin polarization.
If $C_+(s)\neq0, C_-(s)=0$ ( $C_+(s)=0$, $C_-(s)\neq 0$), then
these amplitudes  are exponential functions of time with the
inverse relaxation time $\lambda_+(s)$ ($\lambda_-(s)$) as it can
be seen from Eqs. (\ref{ASolut}-\ref{BSolut}).

Whereas the inverse relaxation time $\lambda_+(s)$ takes its
minimum value at $s=0$ and monotonically increases with parameter
$s$, the inverse relaxation time $\lambda_-(s)$ has a minimum at
 \begin{equation}
 s_m=\sqrt{\frac{C^2}{4D^2}-\frac{\gamma^2}{C^2}}
\label{sm}.
\end{equation}
At this value of $s$,  $\lambda_-(s)$ is equal to
 \begin{equation}
 \lambda_m=3\gamma-\frac{C^2}{4D}-\frac{\gamma^{2}D}{C^2}
\label{lambdam}.
\end{equation}

\begin{figure}[b]
\centering
\includegraphics[width=7.5cm]{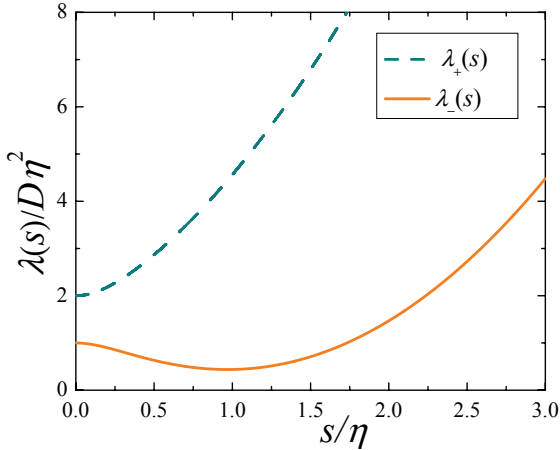}
\caption{(Color online) $s$-dependence of inverse relaxation times
$ \lambda_{\pm}(s)$. The minimum of $\lambda_{-}(s)$  corresponds
to the wave vector giving the longest relaxation time. }
\label{fig7}
\end{figure}

Using relations (\ref{constants}), we find that the minimum value
of $\lambda_-(s)$ is equal to
\begin{equation}
 \lambda_{m}=\frac{7}{16}D\eta^2
\label{lambdam1},
\end{equation}
and it occurs at
\begin{equation}
 s_m=\frac{\sqrt{15}}{4}\eta
\label{sm1}.
\end{equation}
Fig. \ref{fig7} shows $ \lambda_{\pm}(s)$ given by Eq.
(\ref{lambda1}) as a function of $s$.

In order to obtain the general solution of Eqs.
(\ref{SrEq}-\ref{SzrEq}), we should integrate the special
solutions (\ref{SrSpecEq}-\ref{SzrSpecEq}) over positive
parameter $s$ taking into account relations
 (\ref{ASolut}-\ref{lambda}).
 Two arbitrary functions $C_\pm(s)$ entering Eqs. (\ref{ASolut}-\ref{lambda}) can be found from specified initial
 conditions for the radial and $z$-components of polarization
in the form of its Fourier-Bessel transforms
\begin{eqnarray}
 \tilde S_r(s)=\int^{+\infty}_{0}drr S_r(r,0)J_1(sr),\label{SrTilde}\\
 \tilde S_z(s)=\int^{+\infty}_{0}drr S_z(r,0)J_0(sr).
\label{SzTilde}
\end{eqnarray}

As a result of algebraic transformations, we obtain  the solution
of the initial value problem (\ref{SrEq}-\ref{SzrEq}) in the
case of the infinite plane
\begin{eqnarray}
 S_r(r,t)=\int^{+\infty}_{0}dss G_{rr}(r,t,s)\tilde S_r(s)\nonumber\\
+ \int^{+\infty}_{0}dss G_{rz}(r,t,s)\tilde S_z(s),\label{SrSol}\\
 S_z(r,t)=\int^{+\infty}_{0}dss G_{zr}(r,t,s)\tilde S_r(s)\nonumber\\
+ \int^{+\infty}_{0}dss G_{zz}(r,t,s)\tilde S_z(s), \label{SzSol}
\end{eqnarray}
where the Green functions $G_{rr}$, $G_{zr}$, $G_{rz}$, $G_{zz}$
 are defined as follows
\begin{eqnarray}
G_{rr}(r,t,s)=J_1(rs)
\left[\cosh\left(t\sqrt{\gamma^2+C^2s^2}\right)\right.
\nonumber\\
\left.+\gamma\frac{\sinh\left(t\sqrt{\gamma^2+C^2s^2}\right)}
{\sqrt{\gamma^2+C^2s^2}}
 \right]e^{-(Ds^2+3\gamma)t},\label{Grr}
\\
G_{rz}(r,t,s)=-CJ_1(rs)
\frac{\sinh\left(t\sqrt{\gamma^2+C^2s^2}\right)}
{\sqrt{\gamma^2+C^2s^2}}\nonumber\\ \times se^{-(Ds^2+3\gamma)t},
\label{Grz}~\\ G_{zr}(r,t,s)=-CJ_0(rs)
\frac{\sinh\left(t\sqrt{\gamma^2+C^2s^2}\right)}
{\sqrt{\gamma^2+C^2s^2}}\nonumber\\ \times se^{-(Ds^2+3\gamma)t},~
\label{Gzr}\\ G_{zz}(r,t,s)=J_0(rs)
\left[\cosh\left(t\sqrt{\gamma^2+C^2s^2}\right)\right. \nonumber\\
\left.-\gamma\frac{\sinh\left(t\sqrt{\gamma^2+C^2s^2}\right)}
{\sqrt{\gamma^2+C^2s^2}}
 \right]e^{-(Ds^2+3\gamma)t}.~
\label{Gzz}
\end{eqnarray}
Substituting the initial conditions for the radial spin helix
(\ref{Sr}-\ref{Sz}) into Eqs. (\ref{SrTilde},\ref{SzTilde})
and performing integration we find that the generalized functions
$\tilde S_r(s)$ and $\tilde S_z(s)$, which correspond to the
initial conditions (\ref{Sr}) and (\ref{Sz}), act as follows
\begin{eqnarray}
 \int^{+\infty}_{0}dss F(s)\tilde S_r(s)=
 -S_0\frac{d}{dk}\int^{k}_{0}ds\frac{kF(s)}{\sqrt{k^2-s^2}},~\label{SrTildeIC}\\
 \int^{+\infty}_{0}dss F(s)\tilde S_z(s)=
 S_0\frac{d}{dk}\int^{k}_{0}ds\frac{sF(s)}{\sqrt{k^2-s^2}},~\label{SzTildeIC}
\end{eqnarray}
where $F(s)$ is a smooth enough function.

Substituting Eqs. (\ref{SrTildeIC},\ref{SzTildeIC}) into the
general solution (\ref{SrSol},\ref{SzSol}) and taking into
account the expressions for the Green functions
(\ref{Grr}-\ref{Gzz}), we obtain the explicit formulae for the
solution of the drift-diffusion equations
(\ref{SrEq}-\ref{SzrEq}) for the initial conditions
(\ref{Sr}-\ref{Sz}):
\begin{eqnarray}
S_r(r,t)=-S_0\frac{d}{dk}\int_0^k\frac{dsJ_1(sr)}{\sqrt{k^{2}-s^{2}}}\left[
k\cosh\left(t\sqrt{\gamma^2+C^2s^2}\right)\right.\nonumber\\ +
\left.\left(k\gamma+Cs^2 \right)
\frac{\sinh\left(t\sqrt{\gamma^2+C^2s^2}\right)}
{\sqrt{\gamma^2+C^2s^2}}
 \right]e^{-(Ds^2+3\gamma)t},~~~~~~~~ \label{SrSolIC}
\\
S_z(r,t)=S_0\frac{d}{dk}\int_0^k\frac{dssJ_0(sr)}{\sqrt{k^{2}-s^{2}}}\left[
\cosh\left(t\sqrt{\gamma^2+C^2s^2}\right)\right. \nonumber\\
\left.+\left(kC-\gamma \right)
\frac{\sinh\left(t\sqrt{\gamma^2+C^2s^2}\right)}
{\sqrt{\gamma^2+C^2s^2}}
 \right]e^{-(Ds^2+3\gamma)t}.~~~~~~~~~\label{SzSolIC}
\end{eqnarray}

\section{Relaxation of plain spin helix} \label{app:plain}

In the case of the plane spin helix defined by the initial
conditions (\ref{Splnx}-\ref{Splnz}), the solution of the
drift-diffusion equations (\ref{SxEq}-\ref{SzEq}) has the same
harmonical spatial dependence at any moment of time. Therefore, we
can seek the solution of this system of equations in the form
\begin{eqnarray}
  S_x(x,t)=A(k,t)\sin(kx)\label{SxSpecEq},S_y(x,t)=0, \\
  S_z(x,t)=B(k,t)\cos(kx).
 \label{SzSpecEq}
 \end{eqnarray}

Substituting expressions (\ref{SxSpecEq}-\ref{SzSpecEq}) into
the system (\ref{SxEq}-\ref{SzEq}) we obtain the same system  of
ordinary differential equations (\ref{AEq}-\ref{BEq})
 for the functions $A(k,t)$ and $B(k,t)$ as in the axially symmetric case.
Eqs. (\ref{SxSpecEq}-\ref{SzSpecEq}) together with Eqs.
(\ref{ASolut}-\ref{lambda}) determine solutions  of the
drift-diffusion equations (\ref{SxEq}-\ref{SzEq}) with such a
specific harmonic spatial dependence. The arbitrary amplitudes
$C_+$ and $C_-$ are calculated from the initial conditions (Eqs.
(\ref{Splnx}-\ref{Splnz})). Finally, the solution of
drift-diffusion equations (\ref{SxEq}-\ref{SzEq}) for the plane
spin helix is given by
\begin{eqnarray}
S_x(x,t)=-S_0\sin(kx)\left[\cosh\left(t\sqrt{\gamma^2+C^2k^2}\right)\right.\nonumber\\
+ \left.\left(Ck+\gamma\right)
\frac{\sinh\left(t\sqrt{\gamma^2+C^2k^2}\right)}
{\sqrt{\gamma^2+C^2k^2}}
 \right]e^{-(Dk^2+3\gamma)t},~~~~ \label{SxPlane}
\\
S_z(x,t)=S_0\cos(kx)\left[\cosh\left(t\sqrt{\gamma^2+C^2k^2}\right)\right.
\nonumber\\ \left.+\left(Ck-\gamma \right)
\frac{\sinh\left(t\sqrt{\gamma^2+C^2k^2}\right)}
{\sqrt{\gamma^2+C^2k^2}}
 \right]e^{-(Dk^2+3\gamma)t}.~~~~\label{SzPlane}
\end{eqnarray}

\bibliography{spin}

\begin{thebibliography}{23}
\expandafter\ifx\csname natexlab\endcsname\relax\def\natexlab#1{#1}\fi
\expandafter\ifx\csname bibnamefont\endcsname\relax
  \def\bibnamefont#1{#1}\fi
\expandafter\ifx\csname bibfnamefont\endcsname\relax
  \def\bibfnamefont#1{#1}\fi
\expandafter\ifx\csname citenamefont\endcsname\relax
  \def\citenamefont#1{#1}\fi
\expandafter\ifx\csname url\endcsname\relax
  \def\url#1{\texttt{#1}}\fi
\expandafter\ifx\csname urlprefix\endcsname\relax\def\urlprefix{URL }\fi
\providecommand{\bibinfo}[2]{#2}
\providecommand{\eprint}[2][]{\url{#2}}

\bibitem[{\citenamefont{Awschalom et~al.}(2002)\citenamefont{Awschalom,
  Samarth, and Loss}}]{Awschalom02a}
\bibinfo{editor}{\bibfnamefont{D.~D.} \bibnamefont{Awschalom}},
  \bibinfo{editor}{\bibfnamefont{N.}~\bibnamefont{Samarth}}, \bibnamefont{and}
  \bibinfo{editor}{\bibfnamefont{D.}~\bibnamefont{Loss}}, eds.,
  \emph{\bibinfo{title}{Semiconductor Spintronics and Quantum Computation}}
  (\bibinfo{publisher}{Springer-Verlag}, \bibinfo{year}{2002}).

\bibitem[{\citenamefont{Zutic et~al.}(2004)\citenamefont{Zutic, Fabian, and
  Das~Sarma}}]{Zutic04a}
\bibinfo{author}{\bibfnamefont{I.}~\bibnamefont{Zutic}},
  \bibinfo{author}{\bibfnamefont{J.}~\bibnamefont{Fabian}}, \bibnamefont{and}
  \bibinfo{author}{\bibfnamefont{S.}~\bibnamefont{Das~Sarma}},
  \bibinfo{journal}{Rev. Mod. Phys.} \textbf{\bibinfo{volume}{76}},
  \bibinfo{pages}{323} (\bibinfo{year}{2004}).

\bibitem[{\citenamefont{Dyakonov and {Perel'}}(1972)}]{Dyakonov72a}
\bibinfo{author}{\bibfnamefont{M.~I.} \bibnamefont{Dyakonov}} \bibnamefont{and}
  \bibinfo{author}{\bibfnamefont{V.~I.} \bibnamefont{{Perel'}}},
  \bibinfo{journal}{Sov. Phys. Solid State} \textbf{\bibinfo{volume}{13}},
  \bibinfo{pages}{3023} (\bibinfo{year}{1972}).

\bibitem[{\citenamefont{Dyakonov and Kachorovskii}(1986)}]{Dyakonov86a}
\bibinfo{author}{\bibfnamefont{M.~I.} \bibnamefont{Dyakonov}} \bibnamefont{and}
  \bibinfo{author}{\bibfnamefont{V.~Y.} \bibnamefont{Kachorovskii}},
  \bibinfo{journal}{Sov. Phys. Semicond.} \textbf{\bibinfo{volume}{20}},
  \bibinfo{pages}{110} (\bibinfo{year}{1986}).

\bibitem[{\citenamefont{Kiselev and Kim}(2000)}]{Kiselev00a}
\bibinfo{author}{\bibfnamefont{A.~A.} \bibnamefont{Kiselev}} \bibnamefont{and}
  \bibinfo{author}{\bibfnamefont{K.~W.} \bibnamefont{Kim}},
  \bibinfo{journal}{Phys. Rev. B} \textbf{\bibinfo{volume}{61}},
  \bibinfo{pages}{13115} (\bibinfo{year}{2000}).

\bibitem[{\citenamefont{Sherman}(2003)}]{Sherman03a}
\bibinfo{author}{\bibfnamefont{E.~Y.} \bibnamefont{Sherman}},
  \bibinfo{journal}{Appl. Phys. lett} \textbf{\bibinfo{volume}{82}},
  \bibinfo{pages}{209} (\bibinfo{year}{2003}).

\bibitem[{\citenamefont{Weng et~al.}(2004)\citenamefont{Weng, Wu, and
  Shi}}]{Weng04a}
\bibinfo{author}{\bibfnamefont{M.~Q.} \bibnamefont{Weng}},
  \bibinfo{author}{\bibfnamefont{M.~W.} \bibnamefont{Wu}}, \bibnamefont{and}
  \bibinfo{author}{\bibfnamefont{Q.~W.} \bibnamefont{Shi}},
  \bibinfo{journal}{Phys. Rev. B} \textbf{\bibinfo{volume}{69}},
  \bibinfo{pages}{125310} (\bibinfo{year}{2004}).

\bibitem[{\citenamefont{Pershin and Privman}(2004)}]{Pershin04a}
\bibinfo{author}{\bibfnamefont{Y.~V.} \bibnamefont{Pershin}} \bibnamefont{and}
  \bibinfo{author}{\bibfnamefont{V.}~\bibnamefont{Privman}},
  \bibinfo{journal}{Phys. Rev. B} \textbf{\bibinfo{volume}{69}},
  \bibinfo{pages}{073310} (\bibinfo{year}{2004}).

\bibitem[{\citenamefont{Pershin}(2005)}]{Pershin05a}
\bibinfo{author}{\bibfnamefont{Y.~V.} \bibnamefont{Pershin}},
  \bibinfo{journal}{Phys. Rev. B} \textbf{\bibinfo{volume}{71}},
  \bibinfo{pages}{155317} (\bibinfo{year}{2005}).

\bibitem[{\citenamefont{Jiang et~al.}(2005)\citenamefont{Jiang, Weng, Wu, and
  Cheng}}]{Jiang05}
\bibinfo{author}{\bibfnamefont{L.}~\bibnamefont{Jiang}},
  \bibinfo{author}{\bibfnamefont{M.}~\bibnamefont{Weng}},
  \bibinfo{author}{\bibfnamefont{M.}~\bibnamefont{Wu}}, \bibnamefont{and}
  \bibinfo{author}{\bibfnamefont{J.}~\bibnamefont{Cheng}}, \bibinfo{journal}{J.
  Appl. Phys.} \textbf{\bibinfo{volume}{98}}, \bibinfo{pages}{113702}
  (\bibinfo{year}{2005}).

\bibitem[{\citenamefont{Bernevig et~al.}(2006)\citenamefont{Bernevig,
  Orenstein, and Zhang}}]{Bernevig06a}
\bibinfo{author}{\bibfnamefont{B.~A.} \bibnamefont{Bernevig}},
  \bibinfo{author}{\bibfnamefont{J.}~\bibnamefont{Orenstein}},
  \bibnamefont{and} \bibinfo{author}{\bibfnamefont{S.-C.} \bibnamefont{Zhang}},
  \bibinfo{journal}{Phys. Rev. Lett.} \textbf{\bibinfo{volume}{97}},
  \bibinfo{pages}{236601} (\bibinfo{year}{2006}).

\bibitem[{\citenamefont{Schwab et~al.}(2006)\citenamefont{Schwab, Dzierzawa,
  Gorini, and Raimondi}}]{Schwab06a}
\bibinfo{author}{\bibfnamefont{P.}~\bibnamefont{Schwab}},
  \bibinfo{author}{\bibfnamefont{M.}~\bibnamefont{Dzierzawa}},
  \bibinfo{author}{\bibfnamefont{C.}~\bibnamefont{Gorini}}, \bibnamefont{and}
  \bibinfo{author}{\bibfnamefont{R.}~\bibnamefont{Raimondi}},
  \bibinfo{journal}{Phys. Rev. B} \textbf{\bibinfo{volume}{74}},
  \bibinfo{pages}{155316} (\bibinfo{year}{2006}).

\bibitem[{\citenamefont{Koralek et~al.}(2009)\citenamefont{Koralek, Weber,
  Orenstein, Bernevig, Zhang, Mack, and Awschalom}}]{Koralek09a}
\bibinfo{author}{\bibfnamefont{J.~D.} \bibnamefont{Koralek}},
  \bibinfo{author}{\bibfnamefont{C.~P.} \bibnamefont{Weber}},
  \bibinfo{author}{\bibfnamefont{J.}~\bibnamefont{Orenstein}},
  \bibinfo{author}{\bibfnamefont{B.~A.} \bibnamefont{Bernevig}},
  \bibinfo{author}{\bibfnamefont{S.-C.} \bibnamefont{Zhang}},
  \bibinfo{author}{\bibfnamefont{S.}~\bibnamefont{Mack}}, \bibnamefont{and}
  \bibinfo{author}{\bibfnamefont{D.~D.} \bibnamefont{Awschalom}},
  \bibinfo{journal}{Nature} \textbf{\bibinfo{volume}{458}},
  \bibinfo{pages}{610} (\bibinfo{year}{2009}).

\bibitem[{\citenamefont{Kleinert and Bryksin}(2009)}]{Kleinert09a}
\bibinfo{author}{\bibfnamefont{P.}~\bibnamefont{Kleinert}} \bibnamefont{and}
  \bibinfo{author}{\bibfnamefont{V.~V.} \bibnamefont{Bryksin}},
  \bibinfo{journal}{Phys. Rev. B} \textbf{\bibinfo{volume}{79}},
  \bibinfo{pages}{045317} (\bibinfo{year}{2009}).

\bibitem[{\citenamefont{Duckheim et~al.}(2009)\citenamefont{Duckheim, Maslov,
  and Loss}}]{Duckheim09a}
\bibinfo{author}{\bibfnamefont{M.}~\bibnamefont{Duckheim}},
  \bibinfo{author}{\bibfnamefont{D.~L.} \bibnamefont{Maslov}},
  \bibnamefont{and} \bibinfo{author}{\bibfnamefont{D.}~\bibnamefont{Loss}},
  \bibinfo{journal}{Phys. Rev. B} \textbf{\bibinfo{volume}{80}},
  \bibinfo{pages}{235327} (\bibinfo{year}{2009}).

\bibitem[{\citenamefont{Tokatly and Sherman}(2010)}]{Tokatly10a}
\bibinfo{author}{\bibfnamefont{I.~V.} \bibnamefont{Tokatly}} \bibnamefont{and}
  \bibinfo{author}{\bibfnamefont{E.~Y.} \bibnamefont{Sherman}},
  \bibinfo{journal}{Ann. Phys.} \textbf{\bibinfo{volume}{325}},
  \bibinfo{pages}{1104} (\bibinfo{year}{2010}).

\bibitem[{\citenamefont{Bychkov and Rashba}(1984)}]{Bychkov84a}
\bibinfo{author}{\bibfnamefont{Y.}~\bibnamefont{Bychkov}} \bibnamefont{and}
  \bibinfo{author}{\bibfnamefont{E.}~\bibnamefont{Rashba}},
  \bibinfo{journal}{{JETP} Lett.} \textbf{\bibinfo{volume}{39}},
  \bibinfo{pages}{78} (\bibinfo{year}{1984}).

\bibitem[{\citenamefont{Dresselhaus}(1955)}]{Dresselhaus55a}
\bibinfo{author}{\bibfnamefont{G.}~\bibnamefont{Dresselhaus}},
  \bibinfo{journal}{Phys. Rev.} \textbf{\bibinfo{volume}{100}},
  \bibinfo{pages}{580} (\bibinfo{year}{1955}).

\bibitem[{\citenamefont{Weber et~al.}(2007)\citenamefont{Weber, Orenstein,
  Bernevig, Zhang, Stephens, and Awschalom}}]{Weber07a}
\bibinfo{author}{\bibfnamefont{C.~P.} \bibnamefont{Weber}},
  \bibinfo{author}{\bibfnamefont{J.}~\bibnamefont{Orenstein}},
  \bibinfo{author}{\bibfnamefont{B.~A.} \bibnamefont{Bernevig}},
  \bibinfo{author}{\bibfnamefont{S.-C.} \bibnamefont{Zhang}},
  \bibinfo{author}{\bibfnamefont{J.}~\bibnamefont{Stephens}}, \bibnamefont{and}
  \bibinfo{author}{\bibfnamefont{D.~D.} \bibnamefont{Awschalom}},
  \bibinfo{journal}{Phys. Rev. Lett.} \textbf{\bibinfo{volume}{98}},
  \bibinfo{pages}{076604} (\bibinfo{year}{2007}).

\bibitem[{\citenamefont{Saikin et~al.}(2005)\citenamefont{Saikin, Pershin, and
  Privman}}]{Saikin05a}
\bibinfo{author}{\bibfnamefont{S.}~\bibnamefont{Saikin}},
  \bibinfo{author}{\bibfnamefont{Y.}~\bibnamefont{Pershin}}, \bibnamefont{and}
  \bibinfo{author}{\bibfnamefont{V.}~\bibnamefont{Privman}},
  \bibinfo{journal}{{IEE}-Proc. Circ. Dev. Syst.}
  \textbf{\bibinfo{volume}{152}}, \bibinfo{pages}{366} (\bibinfo{year}{2005}).

\bibitem[{\citenamefont{Cameron et~al.}(1996)\citenamefont{Cameron, Riblet, and
  Miller}}]{Cameron96a}
\bibinfo{author}{\bibfnamefont{A.~R.} \bibnamefont{Cameron}},
  \bibinfo{author}{\bibfnamefont{P.}~\bibnamefont{Riblet}}, \bibnamefont{and}
  \bibinfo{author}{\bibfnamefont{A.}~\bibnamefont{Miller}},
  \bibinfo{journal}{Phys. Rev. Lett.} \textbf{\bibinfo{volume}{76}},
  \bibinfo{pages}{4793} (\bibinfo{year}{1996}).

\bibitem[{\citenamefont{Reif}(1965)}]{Reif65a}
\bibinfo{author}{\bibfnamefont{F.}~\bibnamefont{Reif}},
  \emph{\bibinfo{title}{Fundamentals of Statistical and Thermal Physics}}
  (\bibinfo{publisher}{McGraw-Hill}, \bibinfo{year}{1965}).

\bibitem[{\citenamefont{Pershin}(2004)}]{Pershin04b}
\bibinfo{author}{\bibfnamefont{Y.}~\bibnamefont{Pershin}},
  \bibinfo{journal}{Phys. E} \textbf{\bibinfo{volume}{23}},
  \bibinfo{pages}{226} (\bibinfo{year}{2004}).

\end{thebibliography}
\end{document}